\documentclass[prd,showpacs,preprintnumbers,onecolumn,superscriptaddress,amsmath,nofootinbib,amssymb]{revtex4}
\usepackage{graphicx,color,dcolumn,booktabs,bm}
\usepackage{longtable,lscape}
\usepackage{txfonts}
\usepackage{overpic}
\usepackage{amssymb}
\usepackage{epstopdf}
\usepackage{indentfirst}
\usepackage{feynmf}   
\usepackage{slashed}  
\usepackage{cases}
\usepackage{color}
\usepackage{float}
\usepackage{multirow}
\usepackage{graphicx,color,dcolumn,booktabs,bm}
\usepackage{epsfig,dsfont,amssymb,amsmath,amsfonts,amsbsy,mathrsfs}
\usepackage{mathrsfs}

\graphicspath{{Figures/}} %

\usepackage{hyperref}
\hypersetup{colorlinks,citecolor=blue,anchorcolor=red,menucolor=red, linkcolor=red,filecolor=red,runcolor=red,urlcolor=blue,frenchlinks=red}

\makeatletter
\@addtoreset{equation}{section}
\makeatother

\allowdisplaybreaks

\begin{document}

\title{The Excited Charmonium Production in $e^{+}e^{-}$ Annihilation}

\author{Han-Jie Tao}
\affiliation{College of Physics and Electronic Engineering, Northwest Normal University, Lanzhou 730070, China
}
\affiliation{Research Center for Hadron and CSR Physics,
Lanzhou University and Institute of Modern Physics of CAS, Lanzhou 730000, China}
\email{taohanjie1993@163.com}
\author{Yan-Jun Sun}
\email{sunyanjun@mail.nwnu.edu.cn}
\affiliation{College of Physics and Electronic Engineering, Northwest Normal University, Lanzhou 730070, China
}
\affiliation{Research Center for Hadron and CSR Physics, Lanzhou University and Institute of Modern Physics of CAS, Lanzhou 730000, China}
\author{Song-Pei Guo}
\email{1696516235@qq.com}
\affiliation{College of Physics and Electronic Engineering, Northwest Normal University, Lanzhou 730070, China
}
\affiliation{Research Center for Hadron and CSR Physics, Lanzhou University and Institute of Modern Physics of CAS, Lanzhou 730000, China}
\author{Wei Hong}
\affiliation{College of Physics and Electronic Engineering, Northwest Normal University, Lanzhou 730070, China
}
\affiliation{Research Center for Hadron and CSR Physics, Lanzhou University and Institute of Modern Physics of CAS, Lanzhou 730000, China}
\author{Qi Huang}
\affiliation{College of Physics and Electronic Engineering, Northwest Normal University, Lanzhou 730070, China
}
\affiliation{Research Center for Hadron and CSR Physics, Lanzhou University and Institute of Modern Physics of CAS, Lanzhou 730000, China}

\begin{abstract}

We calculate the form factor and cross section of the excited charmonium production process $e^{+}+e^{-}\rightarrow \psi \left ( 2S \right )+\eta _{c}$ by light-cone sum rules. In our method, the form factor depends on the distribution amplitude of $\eta _{c}$ meson. Experimentally, the energy scale of $e^{+}+e^{-}\rightarrow \psi \left ( 2S \right )+\eta _{c}$ process is much larger than the initial energy scale of $\eta _{c}$ meson in our BHL model. Therefore, we further consider the evolution of the distribution amplitude with the energy scale, and select the distribution amplitude as our input parameter when the final effective energy scale is $\mu=5.00\ GeV$. This treatment means that we have chosen the relativistic distribution amplitude. The results show that the relativistic effect contributes greatly to the form factor and cross section. Our results are consistent with Belle experimental data.

\end{abstract}

\pacs{11.55.Hx, 13.66.Bc, 14.40.Lb, 12.39.-x}

\maketitle

\section{introduction}\label{sec1}

The process of double-charmonium production by $e^{+}e^{-}$ annihilation provides us a platform for studying perturbative and non-perturbative effects in quantum chromodynamics (QCD). As early as 2002, the Belle~\cite{Abe:2002rb,Abe:2004ww} experimental group measured the cross section of the exclusive process $e^{+}+e^{-}\rightarrow \psi \left ( 2S \right )+\eta _{c}$,
\begin{eqnarray}
\sigma_{Belle}\times Br_{> 2} =16.3\pm 4.6\pm 3.9\ fb\nonumber.
\end{eqnarray}
Prior to this, the nonrelativistic QCD (NRQCD) and the Color-Singlet model~\cite{Braaten:2002fi} was used to predict the cross-section of the process, but its predictions is much smaller than the experimental results. After the publication of the experimental results, many theoretical works have emerged, such as the method of expanding light cone~\cite{Braguta:2005kr,Braguta:2006nf,Braguta:2012zza}, and the method of perturbative QCD (pQCD), relativistic quark model~\cite{Ebert:2006xq,Ebert:2008kj,Martynenko:2013eoa} and so on~\cite{Liu:2004ga}.

NRQCD factorization is a systematic framework for calculating the production of heavy quarkonium. In Ref.~\cite{Braaten:2002fi}, the NRQCD method was used to predict the cross section of the $e^{+}+e^{-}\rightarrow \psi \left ( 2S \right )+\eta _{c}$ process for the first time, and a result of $\sigma =0.96\pm 0.45\ fb$ smaller than the experimental value by an order of magnitude was acquired, where the leading order contribution and next-leading order (NLO) contribution were taken into account. Obviously, even considering the contribution of the NLO, there was still a big difference between the theoretical and the experimental result. Further, the author took the relativistic correction into consideration and obtained the result of $\sigma =5.6_{-3.3}^{+10.5}\ fb$. It is not hard to see that the relativistic correction has given a considerable contribution. In fact, they considered the contribution of NLO in the past, but the theoretical predictions are still quite different from the experimental result~\cite{Xi-Huai:2014iaa}. The error caused by the NLO is $+10.5$ and $-3.3$ respectively. Thus one would doubt the validity of the expansion methods for the coupling constant $\alpha _{s}$ and $v^{2}$ in the process.

Later, the relativistic effect was particularly considered in the calculation of the
$e^{+}+e^{-}\rightarrow \psi \left ( 2S \right )+\eta _{c}$ cross section. V.V.Braguta \emph{et al}.~\cite{Braguta:2005kr,Braguta:2006nf,Braguta:2012zza} calculated it through the light-cone method. Unlike the NRQCD method, the author of the light cone method adopted a wave function with relativistic effect for the final meson. Their theoretical cross section is $\sigma =16.3\ fb$ for $\psi(2S)$ wave function and $\sigma =10.4_{-7.8}^{+9.2}\ fb$ for $\eta_{c}$ distributed amplitude(DA) , which are in agreement with the experimental result. Besides, based on the perturbative QCD and the relativistic quark model, D.Ebert \emph{et al}~\cite{Ebert:2008kj} had presented a new evaluation for the relativistic effect of double-charmonium. The main improvement lies in an accurate description of the relativistic nature of the  meson wave function. For example, all relativistic contributions of $O\left ( v^{2} \right )$ and $O\left ( v^{4} \right )$ for wave function expansion and relativistic $p/\sqrt{s}$ corrections for the propagation of quark and gluon propagator were considered. The results obtained were also in consistent with the experimental result. This indicates that it is very important to consider its relativistic contribution when calculating the cross section of the $e^{+}+e^{-}\rightarrow \psi \left ( 2S \right )+\eta _{c}$ process.

The method of light-cone sum rules is a fruitful hybrid of the SVZ sum rules~\cite{Shifman:1978bx,Colangelo:2000dp} technique and the theory of hard exclusive processes, it is be a powerful tool for calculating the form factors in the large momentum transfer. In our previous work, we have successfully applied this method to the $e^{+}+e^{-}\rightarrow J/\psi +\eta _{c}$ process~\cite{Sun:2009zk}.
In the present study, we will extend this method to the new process $e^{+}+e^{-}\rightarrow \psi \left ( 2S \right )+\eta _{c}$. Unlike
the particle $J/\psi$ which is a ground state meson, $\psi \left ( 2S \right )$ here is a excited state meson. However, after careful treatment for the hadronic spectra, we find that that the light-cone sum rules can also be used for this process. In this paper, firstly, we give the form factor within LCSR. Secondly, we discuss the evolution of the distribution amplitude(DA), which is an input parameter for the form factor, with the energy scales. Thirdly, we get the results of form factor and cross section and compare our results with that of Belle experiment and other theoretical methods. The last section is a summary.

\section{the $e^{+}+e^{-}\rightarrow \psi \left ( 2S \right )+\eta _{c}$ form factor within the LCSR}\label{sec2}

The start object of light-cone sum rules is a T-product of two quark currents sandwiched between vacuum and on-shell state. For the exclusive process $e^{+}\left ( P_{1} \right )e^{-}\left ( P_{2} \right )\rightarrow\psi \left ( 2S \right )\left ( P_{3} \right )\eta _{c}\left ( P_{4} \right )$, the corresponding amplitude has the structure of a typical LCSR correlation function
\begin{eqnarray}
\label{eq9}
\Pi_{\mu \nu }\left ( P_{4},q \right )=i\int d^{^{4}}xe^{-iqx}< \eta _{c}\left ( P_{4} \right )\mid T\left \{ J_{\mu }^{c}\left ( x \right )J_{\nu }^{c}\left ( 0 \right ) \right \}\mid 0>,
\end{eqnarray}
with $J_{\mu }^{c}=\bar{C}\left ( x \right )\gamma _{\mu }C\left ( x \right )$ is the colorless c-quark electromagnetic current, and $q$ is the four-momentum of the virtual photon, $P_{4}$ stands for the four-momentum of $\eta _{c}$ meson.

Generally, we have two ways to dealing with the correlation function. One is to insert a complete states into it, another is to expand the operator product(OPE) in it. Firstly, we insert a complete intermediate states with the same quantum number as the $J/\psi $ ground state and its first excited state $\psi \left ( 2S \right )$ between the two electromagnetic currents. Then, the contributions of the $J/\psi $ ground state, the first excited state $\psi \left ( 2S \right )$ and the higher excited and continuum states can be separated. The simplified time-like hadronic states representation is
\begin{eqnarray}
\label{eq1}
\Pi _{\mu \nu }\left ( P_{4},q \right )&=&\frac{1}{m_{J/\psi }^{2}-\left ( q-P_{4} \right )^{2}}<\eta _{c}\left ( P_{4} \right ) \mid J_{\mu }^{c}\left ( 0 \right )\mid  J/\psi \left ( P_{4} -q\right )\mid 0> < J/\psi \left (P_{4} -q\right )\mid J_{\nu }^{c}\left ( 0 \right )\mid 0>\nonumber\\
&&+\frac{1}{m_{\psi \left ( 2S \right ) }^{2}-\left ( q-P_{4} \right )^{2}}<\eta _{c}\left ( P_{4} \right ) \mid J_{\mu }^{c}\left ( 0 \right )\mid \psi \left ( 2S \right ) \left ( P_{4} -q\right )\mid 0>< \psi \left ( 2S \right )\left (P_{4} -q\right )\mid J_{\nu }^{c}\left ( 0 \right )\mid 0>\nonumber\\
&&+\frac{1}{\pi }\int_{s_{02}}^{\infty }ds\frac{\rm{Im}\Pi _{\mu \nu }}{s-\left ( q-P_{4} \right )^{2}},
\end{eqnarray}
where $m_{J/\psi }$ is the $J/\psi $ mass, $m_{\psi \left ( 2S \right )}$ is the mass of $\psi \left ( 2S \right )$. The first and second terms represent the contributions from $J/\psi $ and $\psi \left ( 2S \right )$, respectively. And the third one is the dispersion integral that includes the contributions from the higher excited and continuum states.
The matrix elements~\cite{Sun:2009zk} in Eq.(\ref{eq1}) are  defined as
\begin{eqnarray}
\label{eq2}
< 0\mid J_{\nu  }^{c}\left ( 0 \right )\mid J/\psi \left ( P_{4}-q \right )>&=&f_{J/\psi }m_{J/\psi }\epsilon _{\nu }^{{}'},\nonumber\\
< 0\mid J_{\nu  }^{c}\left ( 0 \right )\mid \psi \left ( 2S \right ) \left ( P_{4}-q \right )>&=&f_{\psi \left ( 2S \right ) }m_{\psi \left ( 2S \right ) }\epsilon _{\nu },\nonumber\\
< \eta_{c}\left ( P_{4} \right )\mid J_{\mu }^{c}\left ( 0 \right )\mid J/\psi \left ( P_{4}-q \right )> &=&\epsilon _{\mu abc}\epsilon {}'^{a\ast }q^{b}P_{4}^{c}F_{VP}^{{}'},\nonumber\\
< \eta _{c}\left ( P_{4} \right )\mid J_{\mu }^{c}\left ( 0 \right )\mid \psi \left ( 2S \right ) \left (P_{4}-q \right )>&=&\epsilon _{\mu abc}\epsilon ^{a\ast }q^{b}P_{4}^{c}F_{VP},
\end{eqnarray}
where $F_{VP}^{{}'}$ and $F_{VP}$ are the form factors of $e^{+}e^{-}\rightarrow J/\psi \eta _{c}$ and $e^{+}e^{-}\rightarrow \psi \left ( 2S \right ) \eta _{c}$ respectively, $f_{J/\psi }$ and $f_{\psi \left ( 2S \right ) }$ represent the decay constants of $J/\psi $ and $\psi \left ( 2S \right )$ severally, $\epsilon _{\nu }^{{}'}$ and $\epsilon _{\nu }$ are the polarization vector of $J/\psi $ and $\psi \left ( 2S \right )$ respectively.
With these definition of matrix elements, the hadronic state representation is converted into the form
\begin{eqnarray}
\label{eq3}
\Pi _{\mu \nu }\left ( P_{4},q \right )&=-\epsilon _{\mu \nu \alpha \beta }q^{\alpha }P_{4}^{\beta }\left [ \frac{F_{VP}{}'f_{J/\psi }}{m_{J/\psi }^{2}-\left ( q-P_{4} \right )^{2}}\right.
\left.+\frac{F_{VP}f_{\psi \left ( 2S \right )}}{m_{\psi \left ( 2S \right )}^{2}-\left ( q-P_{4} \right )^{2}}\right ]+\frac{1}{\pi }\int_{s_{02}}^{\infty }ds\frac{{\rm{Im}}\Pi _{\mu \nu }\left (P_{4} ,s\right )}{s-\left ( q-P_{4} \right )^{2}}.
\end{eqnarray}
Note that the first is the contribution from the ground state hadronic spectra, while the second is the contribution from the first excited hadronic spectra. This is very different from the usual light-cone sum rules where only ground state contribution is separated. However, this can be done since we just separate more states from the complete hadronic states.

At sufficiently large $-q^{2}=Q^{2}=s$ and $\mid (P_{4}-q)^{2}\mid$, the dominant part of the integrand in the correlation function Eq.(\ref{eq9}) stems from the region near the light-cone $x^{2}=0$. Thus, we calculate the leading order contribution to the light-cone OPE of the correlator. Contracting the c-quark fields in Eq.(\ref{eq9}), for simplicity, we adopt the free c-quark propagator
\begin{eqnarray}
iS\left ( x,0 \right )=i\int\frac{d^{4}k}{\left ( 2\pi  \right )^{4}}e^{-ikx}\frac{\not{k}+m_{c}}{k^{2}-m_{c}^{2}},
\end{eqnarray}
where $m_{c}$ and $k$ are the mass and the four-momentum of c-quark. For the subsequent matrix elements, we parameterize it with the integral of the distribution amplitudes of the $\eta _{c}$ meson
\begin{eqnarray}
< \eta _{c}\left ( P_{4} \right )\mid \bar{C}\left ( x \right )\gamma ^{\beta }\gamma _{5}C\left ( 0 \right )\mid 0> &=-iP_{4}^{\beta }f_{\eta _{c}}\int_{0}^{1}due^{iuP_{4}x}\phi_{\eta _{c}}\left ( u \right )+higher\ twist\ terms,
\end{eqnarray}
where $u$ is longitudinal momentum rate, $f_{\eta _{c}}$ is the dacay constant of $\eta _{c}$ meson, and $\phi _{\eta _{c}}$ is the $\eta _{c}$ light-cone DA. The final expression of the operator product expansion for the correlation function is
\begin{eqnarray}
\label{eq6}
\Pi _{\mu \nu }\left ( P_{4} ,q\right )=2\epsilon _{\mu \nu \alpha \beta }q^{\alpha }P_{4}^{\beta }f_{\eta _{c}}\int_{0}^{1}dx\frac{\phi _{\eta _{c}}\left ( x \right )}{m_{c}^{2}-\left ( xP_{4}-q \right )^{2}}.
\end{eqnarray}

However, the third term in Eq.(\ref{eq3}), which corresponds to higher excited and continuum states is not clear. Since at $q^{2}\rightarrow -\infty$, the limit $\Pi _{\mu \nu }\left ( P_{4},q \right )\rightarrow \Pi _{\mu \nu }^{(pert)}\left ( P_{4},q \right )$ is valid, we convert the integrand in Eq.(\ref{eq3}) into another form, called quark-hadron duality
\begin{eqnarray}
\frac{1}{\pi }{\rm{Im}}\Pi_{\mu \nu }^{(pert)}\left ( P_{4},s \right )=&2\epsilon _{\mu \nu \alpha \beta }f_{\eta _{c}}\int_{0}^{1}dx\phi _{\eta _{c}}\left ( x \right )\delta \left ( m_{c}^{2}\right.
\left.+x\bar{x}P_{4}^{2}-\bar{x}q^{2}-\bar{x}s_{02} \right ),
\end{eqnarray}
with $\bar{x}=1-x$ and
\begin{eqnarray}
\frac{1}{\pi }\int_{s_{02}}^{\infty }ds\frac{{\rm{Im}}\Pi _{\mu \nu }\left ( P_{4},s \right )}{s-\left ( q-P_{4} \right )^{2}}\simeq \frac{1}{\pi }\int_{s_{02}}^{\infty }ds\frac{{\rm{Im}}\Pi _{\mu \nu }^{(pret)}\left ( P_{4},s \right )}{s-\left ( q-P_{4} \right )^{2}}\simeq 2\epsilon _{\mu \nu \alpha \beta }f_{\eta _{c}}\int_{0}^{\Delta _{2}}dx\frac{\phi _{\eta _{c}}\left ( x \right )}{m_{c}^{2}-\left ( xP_{4}-q \right )^{2}}.
\end{eqnarray}

Finally, the hadronic states representation in Eq.(\ref{eq3}) is matched with OPE in Eq.(\ref{eq6}). In order to suppresses the third term in Eq.(\ref{eq3}), Borel transformation is applied to both the hadronic side and the OPE side~\cite{Shifman:1978bx,Shifman:1978by}
\begin{eqnarray}
B_{M^{2}}\frac{1}{m_{J/\psi }^{2}-\left ( q-P_{4} \right )^{2}}&=&\frac{1}{M^{2}}e^{-\frac{m_{J/\psi }^{2}}{M^{2}}},\nonumber\\
B_{M^{2}}\frac{1}{m_{\psi \left ( 2S \right ) }^{2}-\left ( q-P_{4} \right )^{2}}&=&\frac{1}{M^{2}}e^{-\frac{m_{\psi \left ( 2S \right ) }^{2}}{M^{2}}},\nonumber\\
B_{M^{2}}\frac{1}{m_{c }^{2}-\left ( q-xP_{4} \right )^{2}}&=&\frac{1}{xM^{2}}e^{\left \{ -\frac{1}{xM^{2}}\left [ m_{c }^{2}+x\left ( 1-x \right )P_{4}^{2}-\left ( 1-x \right )q^{2} \right ] \right \}},
\end{eqnarray}
where $M^{2}$ is the Borel parameter. After the Borel transformation, the light-cone sum rule for form factor is obtained
\begin{eqnarray}
\label{eq5}
F_{VP}=\frac{2f_{\eta _{c}}}{m_{\psi \left ( 2S \right )}f_{\psi \left ( 2S \right )}}\int_{\Delta _{2}}^{\Delta _{1}}dx\frac{\phi _{\eta _{c}\left ( x \right )}}{x}e^{\left \{ -\frac{1}{xM^{2}}\left [ m_{c }^{2}+x\left ( 1-x \right )m_{\eta_{c}}^{2}-\left ( 1-x \right )q^{2} \right ] +\frac{m_{\psi {(2S)}}^{2}}{M^{2}}\right \}},
\end{eqnarray}
with
\begin{eqnarray}
&\Delta _{1}=\frac{1}{2m_{\eta _{c}}^{2}}\left [ \sqrt{\left ( s_{01}-m_{\eta _{c}}^{2}+Q^{2} \right )^{2}+4\left ( m_{c}^{2}+ Q^{2}\right )m_{\eta _{c}}^{2}}-\left ( s_{01}-m_{\eta _{c}}^{2}-q^{2} \right ) \right ],\nonumber\\&
\Delta _{2}=\frac{1}{2m_{\eta _{c}}^{2}}\left [ \sqrt{\left ( s_{02}-m_{\eta _{c}}^{2}+Q^{2} \right )^{2}+4\left ( m_{c}^{2}+ Q^{2}\right )m_{\eta _{c}}^{2}}-\left ( s_{02}-m_{\eta _{c}}^{2}-q^{2} \right ) \right ],
\end{eqnarray}
where $s_{01}$ and $s_{02}$ are the first and excited state threshold parameters, $-q^{2}=Q^{2}=s$~\cite{Sun:2009zk}. Actually, apart from that, the expression of the form factor of the process $e^{+}+e^{-}\rightarrow \psi \left ( 2S \right )+\eta _{c}$ is exactly the same as that of the process $e^{+}+e^{-}\rightarrow J/\psi \left ( 2S \right )+\eta _{c}$~\cite{Sun:2009zk}. To see that, it is only necessary to change the mass and decay constant of the $J/\psi$ in the literature~\cite{Sun:2009zk} to that of the $\psi(2S)$, and to change the integral interval of $\Delta _{1}\sim 1$ to that of $\Delta _{2}\sim\Delta _{1}$ here. Generally speaking, the above method of deriving the form factor is also applicable to the process of hadronic exclusive production by $e^{+}e^{-}$ annihilation with large momentum transfer. It is found that the form factor depends on the $\eta _{c}$ meson light cone DA, especially on its end point behavior due to $\Delta _{2}\simeq 0.88$ to $\Delta _{1}\simeq 0.90$.

\section{light-cone DA of $\eta _{c}$ meson}\label{sec3}

At present, it is difficult to give the light-cone wave function from the first principle of QCD. Thus one usually constructs some phenomenological models for the wave function, for instance BHL model~\cite{BHL}, BLL model~\cite{Braguta:2006wr,Braguta:2012zza}, BC model~\cite{Bondar:2004sv}, BKL model~\cite{Bodwin:2006dm}, MS model~\cite{Ma:2004qf} and etc. The key input for the form factor is the gauge-independent and process-independent DA $\phi _{\eta _{c}}\left ( x \right )$, which is of non-perturbative nature and can be defined as the integral of the valence Fock wave function~\cite{Lepage:1980fj}
\begin{eqnarray}
\label{eq7}
\phi _{\eta _{c}}\left ( x ,\mu _{0}\right )=\frac{2\sqrt{6}}{f_{\eta _{c}}}\int_{\left | \vec{k_{\perp }} \right |^{2}<\mu_{0}^{2}}\frac{d^{2}\vec{k_{\perp }}}{16\pi ^{3}}\Psi _{\eta _{c}}\left ( x,\vec{k_{\perp }}\right ),
\end{eqnarray}
$\mu _{0}$ stands for the separation scale between the perturbative and non-perturbative regions. For the massive quark-antiquark system, Ref.~\cite{H} provides a good solution $\varphi _{BHL}( x,\vec{k}_{\perp })=Ae^{-b^{2}\frac{\vec{k}_{\perp }^{2}+m_{c}^{\ast2}}{x(1-x)}}$ of the bound state by solving the Bethe-Salpeter equation with the harmonic oscillator potential in the instantaneous approximation. Furthermore, the spin structure of the light-cone wave function should be connected with that of the instant-form wave function by considering the Wigner-Melosh rotation~\cite{Huang:2007kb},
\begin{eqnarray}
\Psi _{\eta _{c}}^{\lambda _{1}\lambda _{2}}\left ( x,\vec{k}_{\perp } \right )=\varphi _{BHL}( x,\vec{k}_{\perp })\chi^{\lambda_{1} \lambda _{2}}( x,\vec{k}_{\perp })=Ae^{-b^{2}\frac{\vec{k}_{\perp }^{2}+m_{c}^{\ast2}}{x(1-x)}}\chi^{\lambda _{1}\lambda _{2}}( x,\vec{k}_{\perp }),
\end{eqnarray}
where $m_{c}^{\ast}$ denotes the constituent mass of c-quarks, $\chi^{\lambda_{1} \lambda _{2}}( x,\vec{k}_{\perp })=\frac{Am_{c}^{\ast }}{\sqrt{\vec{k}_{\perp }^{2}+m_{c}^{\ast 2}}}$ is an spin-space wave function~\cite{Melosh:1974cu}, $\lambda _{1}$ and $\lambda _{2}$ stand for the helicity of the constitute $c$ and $\bar{c}$. The parameters $A$ and $b^{2}$ can be restricted by two constraints on them absolutely. One constraint is from the relativistic wave function normalization~\cite{Huang:2007kb}
\begin{eqnarray}
\label{eq10}
\frac{2\sqrt{6}}{f_{\eta _{c}}}\int_{0}^{1}dx\int_{\left | \vec{k_{\perp }} \right |^{2}<\mu_{0}^{2}}\frac{d^{2}\vec{k_{\perp }}}{16\pi ^{3}}\sum\Psi _{\eta _{c}}\left ( x,\vec{k_{\perp }}\right )=1,
\end{eqnarray}
another one from the probability of finding the $\mid c\bar{c}>$ Fock state in the charmonium,
\begin{eqnarray}
\label{eq11}
\int_{0}^{1}dx\int\frac{d^{2}\vec{k_{\perp }}}{16\pi ^{3}}|\varphi _{BHL}\left ( x,\vec{k}_{\perp } \right )|^{2}=P_{\eta _{c}},
\end{eqnarray}
with $P_{\eta _{c}}\simeq 0.8$~\cite{Huang:1994dy}, where we adopt the initial energy $\mu _{0}=m_{c}^{\ast }=1.8\ GeV$~\cite{Sun:2009zk}. With the decay constant $f_{\eta _{c}}=0.335\ GeV$~\cite{Tanabashi:2018oca}, we obtain two parameters $A=285.64\ GeV^{-1}$ and $b^{2}=0.19057\ GeV^{-2}$. To sum up, we can determine the DA of $\eta_{c}$ meson at initial scale $\mu_{0}$, with the parameters $A$ and $b^{2}$ in the wave function constrained by Eq.(\ref{eq10}) and Eq.(\ref{eq11}).

From Eq.(\ref{eq7}), the distribution amplitude relies upon the energy scale $\mu $ also. As for a scale $\mu>\mu_{0}$, the non-perturbative DA is given by Eq.(\ref{eq7}) or explicitly by the renormalization group evolution~\cite{Lepage:1980fj}. The solution are Gegenbauer polynomials expansions as
\begin{eqnarray}
\phi _{\eta _{c}}\left ( x_{i},\mu  \right )=x_{1}x_{2}\sum_{n=0}^{\infty }a_{n}\left ({\rm{ln}}\frac{\mu ^{2}}{\Lambda ^{2}} \right )^{-\gamma _{n}}C_{n}^{\frac{2}{3}}\left ( x_{1}-x_{2} \right ),
\end{eqnarray}
with
\begin{eqnarray}
\gamma _{n}=\frac{C_{F}}{\beta }\left ( 1+4\sum_{k=2}^{n+1}\frac{1}{k}-\frac{2\delta _{h_{1}\bar{h_{2}}}}{\left ( n+1 \right )\left ( n+2 \right )} \right )\geq 0
\end{eqnarray}
is the non-singlet anomalous dimensions,
\begin{figure}[!htbp]
\centering
\begin{tabular}{c}
\includegraphics[width=0.5\textwidth]{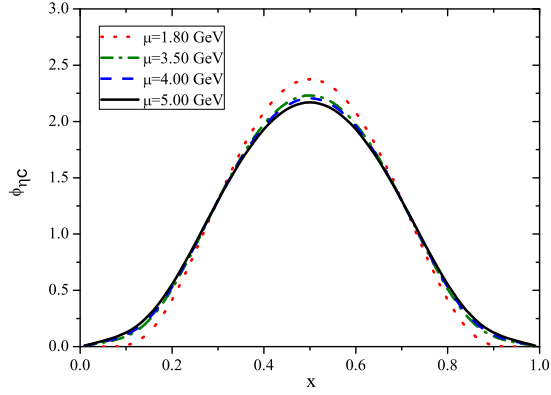}
\end{tabular}
\caption{(color online) Distribution amplitude of $ \eta _{c}$ meson in the BHL model~\cite{BHL} at some typical energy scales. The dotted line, dash-dotted line, dashed line and solid line represent DA at $\mu=1.8\ GeV$, $\mu=3.50\ GeV$, $\mu=4.00\ GeV$ and $\mu=5.00\ GeV$ respectively.}\label{potentialshape}
\end{figure}
and $C_{n}^{\frac{2}{3}}$ is the Gegenbauer functions. $C_{F}=4/3$, $\delta _{h_{1}\bar{h_{2}}}=1$ when the $c$ and $\bar{c}$ helicities are opposite. Obviously, the non-perturbative coefficients $a_{n}$ depends on the scales. At $\mu \rightarrow \infty $, $a_{n}$ vanish, and the limit $a_{n}=0$ corresponds to the asymptotic distribution amplitude $\phi_{as}\left ( x \right )=6x\left ( 1-x \right )$~\cite{Colangelo:2000dp}.
 FIG.1 shows the dependence of the DA on the energy scales. It can be seen from the figure that with the increase of the energy scale, the middle of the $\phi _{\eta _{c}}\left ( x_{i},\mu  \right )$ decreases and the two ends of it increase. Further, substituting the DA of $\eta _{c}$ into Eq.(\ref{eq5}), we find the form factor increasing with the increase of the energy scale, since the integral interval of the form factor is $\Delta _{2}\simeq 0.88$ to $\Delta _{1}\simeq 0.90$. We can better understand this from a physical point of view. As the energy scale gets higher and higher, high-energy tail of the DA becomes larger and larger, meaning an increasing relativity of the DA. Therefore, the relativistic effect contained in the corresponding form factor is also increasing.
\section{cross section}

Generlly, the cross section of $e^{+}\left ( P_{1} \right )e^{-}\left ( P_{2} \right )\rightarrow\psi \left ( 2S \right )\left ( P_{3} \right )\eta _{c}\left ( P_{4} \right )$ can be expressed as
\begin{eqnarray}
\label{eq8}
\sigma =\frac{1}{4E_{1}E_{2}\left | v_{r} \right |}\int \frac{d^{3}\vec{P}_{3}d^{3}\vec{P_{4}}}{\left ( 2\pi  \right )^{3}2E_{1}\left ( 2\pi  \right )^{3}2E_{2}}\left ( 2\pi  \right )^{4}\delta ^{4}\left ( P_{1}+P_{2}-P_{3}-P_{4} \right )\left | \overline{\mathcal{M}}\right |^{2},
\end{eqnarray}
where $| v_{r}|$ is the relative velocity of the initial positive and negative electrons, $\left |\overline{\mathcal{M}}\right |^{2}$ is related to scattering matrix element $\mathcal{M}$ as
\begin{eqnarray}
\left |\overline{\mathcal{M}}\right |^{2}=\frac{1}{4}\sum_{s,t}\left |\mathcal{M}  \right |^{2},
\end{eqnarray}
with the scattering matrix element~\cite{Sun:2009zk}
\begin{eqnarray}
\mathcal{M}=i\int d^{4}x< \psi \left ( 2S \right )\eta _{c}\mid T\left \{ Q_{c}J_{\mu }^{c}\left ( x \right )A^{\mu }\left ( x \right ), Q_{e}\bar{\psi }\left ( 0 \right )\gamma _{\nu }\psi \left ( 0 \right )A^{\nu }\left ( 0 \right ) \right \}\mid e^{+}e^{-}>.
\end{eqnarray}
After some tedious calculations, we get
\begin{eqnarray}
\left |\overline{\mathcal{M}}\right |^{2}=\frac{Q_{c}^{2}Q_{e}^{2}}{8s}\left | F_{VP} \right |^{2}\left [ s-\left ( m_{\psi \left ( 2s \right )}+m_{\eta _{c}} \right )^{2} \right ]\left [ s-\left ( m_{\psi \left ( 2s \right )}-m_{\eta _{c}} \right )^{2} \right ]\left ( \rm{cos}^{2}\theta -1 \right ),
\end{eqnarray}
where $Q_{c}$ and $Q_{e}$ denote the charge of c-quark and electron, respectively, $\theta $ is the scattering angle, and $s$ is the invariant mass square of the $e^{+}e^{-}$ system. $F_{VP}$ is the form factor of $e^{+}e^{-}\rightarrow \psi \left ( 2S \right ) \eta _{c}$ as in section II. In order to compare with the experiment, we adopt the invariant mass square $s=112\ GeV^{2}$. Substitute the $\left |\overline{\mathcal{M}}\right |^{2}$ into Eq.(\ref{eq8}), integrate the phase space, at last, we obtain
\begin{eqnarray}
\sigma =\frac{\alpha ^{2}Q_{c}^{2}\pi }{6s^{3}}\left \{ \left [ s-\left ( m_{\psi \left ( 2S \right )}+m_{\eta _{c}} \right )^{2} \right ]\left [ s-\left ( m_{\psi \left ( 2s \right )}-m_{\eta _{c}} \right )^{2} \right ] \right \}^{\frac{2}{3}}\left | F_{VP} \right |^{2}.
\end{eqnarray}
In section II and III, we show that the form factor is the function of DA, which depending on the energy scale. Thus, we will discuss the form factor and cross section numerically.

\section{Numerical results and discussion}\label{sec4}

In order to obtain the numerical results of the form factor and cross section, several parameters are taken as $m_{c}=1.27\ GeV$,  $m_{\eta _{c}}=2.979\ GeV$, $m_{\psi \left ( 2S \right )}=3.686\ GeV$, $f_{\eta _{c}}=0.416\ GeV$ and $f_{\psi \left ( 2S \right )}=0.295\ GeV$~\cite{Tanabashi:2018oca,Edwards:2000bb,Braga:2015jca,Negash:2015rua}.
\begin{figure}[!htbp]
\centering
\begin{tabular}{c}
\includegraphics[width=0.5\textwidth]{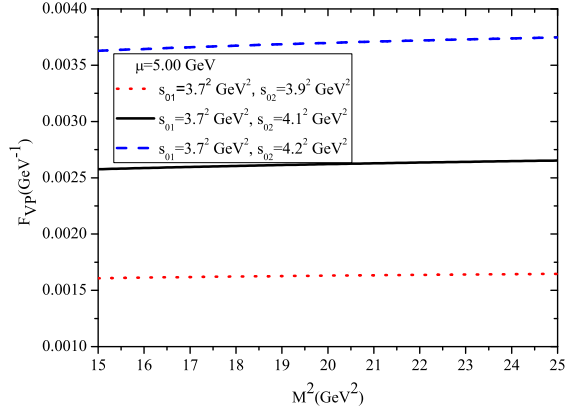}
\end{tabular}
\caption{(color online) The dependence of form factor on the threshold parameter $s_{02}$ within the LCSR approach. Where the energy scale is $\mu=5.00\ GeV $ and the threshold parameter $s_{01}=3.7^{2}\ GeV^{2}$. }\label{potentialshape}
\end{figure}

In our methods, there are two thresholds in the form factor Eq.(\ref{eq5}), let's first look at the dependence of the results on the thresholds. In order to see this point, we change each threshold  within a certain range, $3.6^{2}\ GeV^{2}<s_{01}<3.8^{2}\ GeV^{2}$ and $3.9^{2}\ GeV^{2}<s_{02}<4.2^{2}\ GeV^{2}$, as can be seen from FIG.2. One thing that needs to be added, in our previous work, we had taken threshold as $3.7^{2}\ GeV^{2}<s_{02}<4.1^{2}\ GeV^{2}$ for the $e^{+}+e^{-}\rightarrow J/\psi +\eta _{c}$ process~\cite{Sun:2009zk,Eidemuller:2000rc}, but in the current situation, there are two different thresholds, so we do a slightly different choice. Since our results depend a little on threshold, the central thresholds parameters we determined here are $s_{01}=3.7^{2}\ GeV^{2}$, $s_{02}=4.1^{2}\ GeV^{2}$. Our results do not depend on the Borel parameter, which is taken as $15\ GeV^{2}\leq M^{2}\leq 25\ GeV^{2}$.
\begin{figure}[!htbp]
\centering
\begin{tabular}{c}
\includegraphics[width=0.5\textwidth]{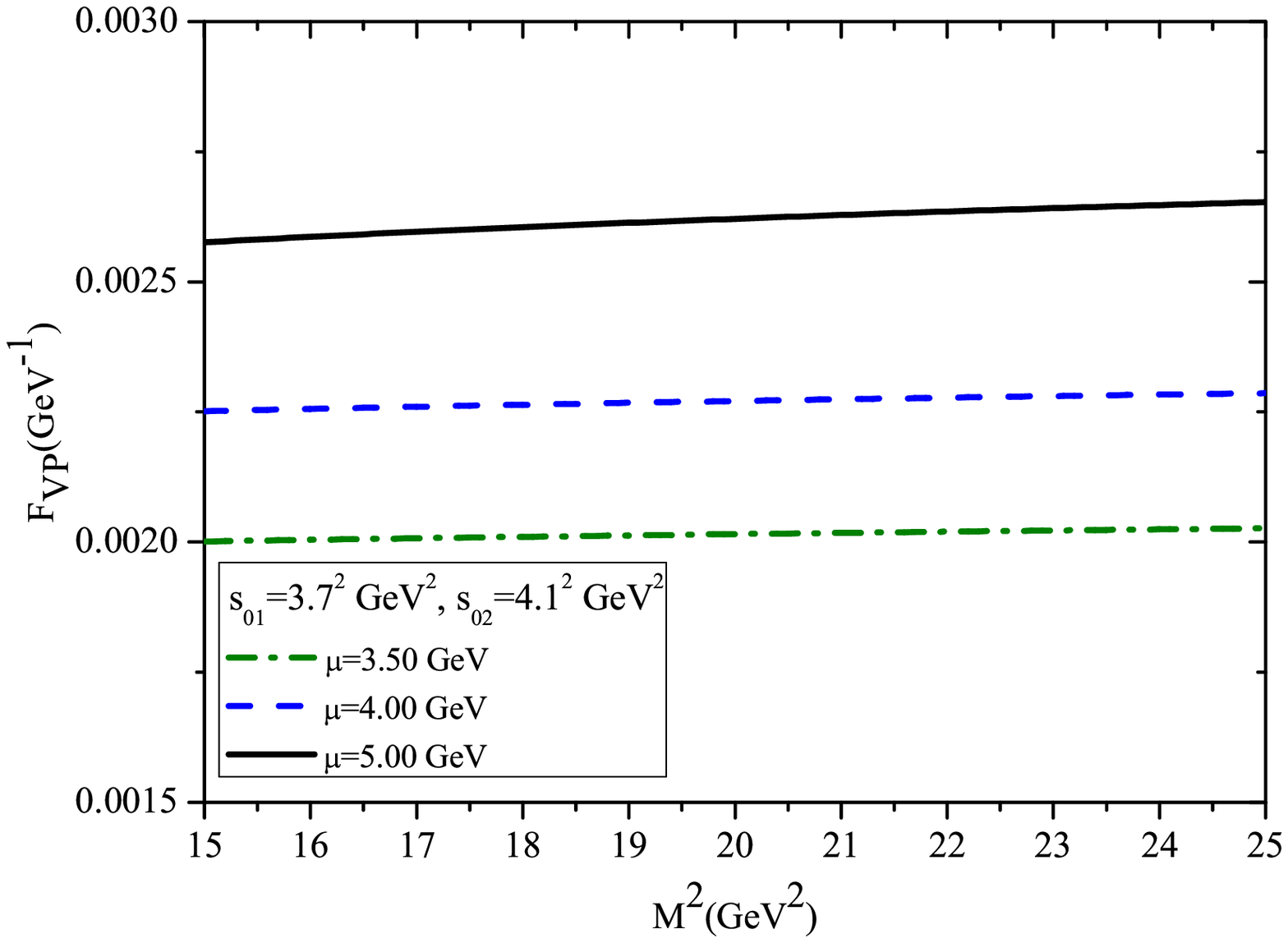}
\includegraphics[width=0.5\textwidth]{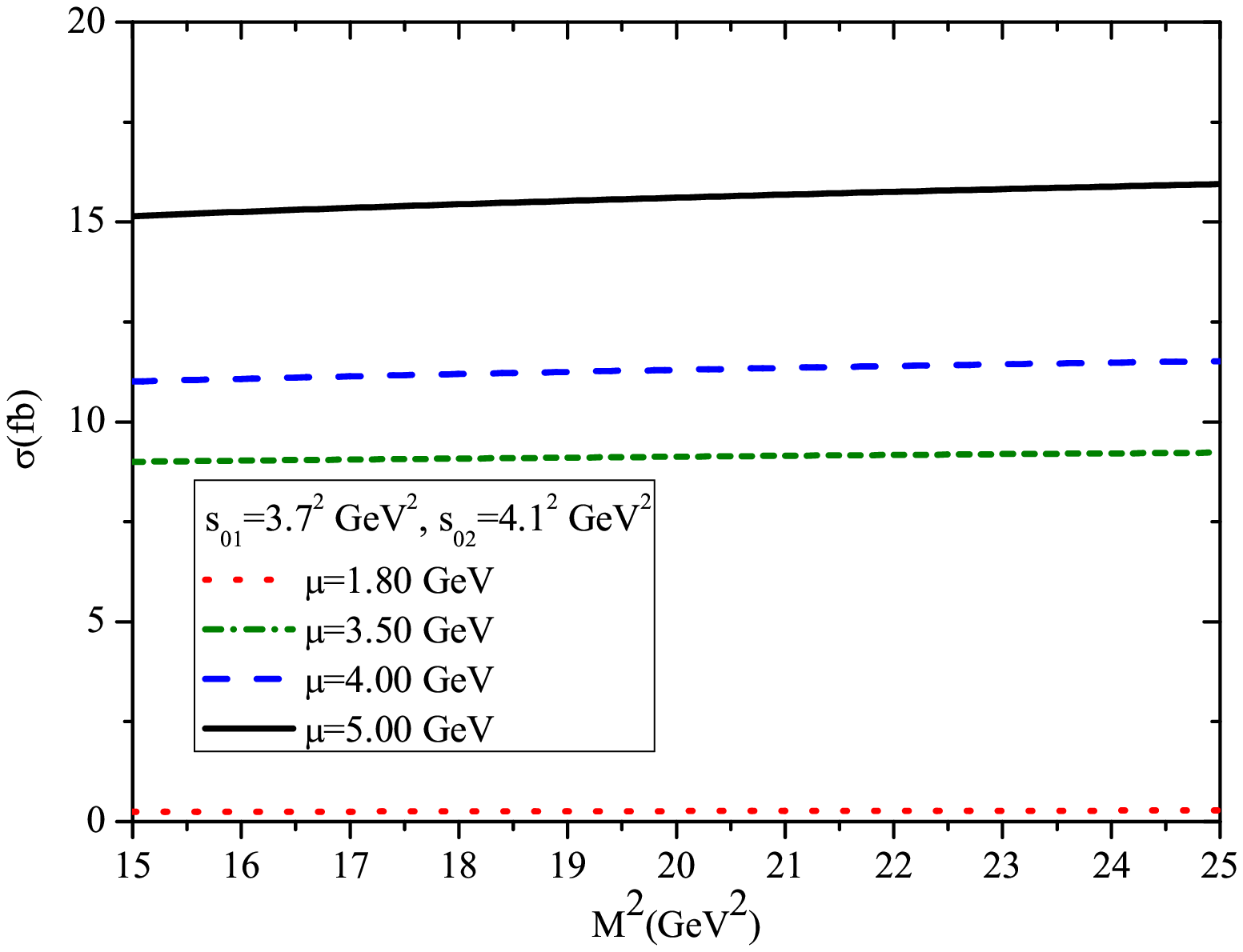}\\%
\end{tabular}
\caption{(color online) The dependence of form factor and cross section on energy scale in the LCSR approach. The left corresponds to the scale $\mu=3.50\ GeV$ (dash-dotted line), $\mu=4.00\ GeV$ (dashed line), $\mu=5.00\ GeV$ (solid line), and the right corresponds to the scale $\mu=1.80\ GeV$ (dotted line), $\mu=3.50\ GeV$ (dash-dotted line), $\mu=4.00\ GeV$ ( dashed line), $\mu=5.00\ GeV$ (solid line). Where the threshold parameters are $s_{01}=3.7^{2}\ GeV^{2}$ and $s_{02}=4.1^{2}\ GeV^{2}$. }\label{potentialshape}
\end{figure}

FIG.3 shows the dependence of form factor and cross section on the energy scale $\mu$. As can be seen from the figure, with the increasing of energy scale, the form factor and cross section become larger and larger. Since Eq.(\ref{eq5}) indicates the form factor is an integral in the interval $0.88\leq x \leq0.90$, while the factor other than the distribution amplitude in this interval is nearly a constant, so the form factor depends only on DA. When the energy scale increases, FIG.1 shows that the high-energy tail of DA is getting bigger and bigger, thus, the form factor becomes larger and larger accordingly. Similar trends are observed for cross section. As the high-energy tail of DA means a large relativity of the DA, the relativistic effect contained in the corresponding form factor and cross section is also large. We show this clearly in TABLE I. As for the effective scale $\mu$ of the process, Ref~\cite{Bondar:2004sv,Sun:2009zk} suggested $\mu \approx 3.5\ GeV$, while another usually adopted scale is $\mu=\frac{\sqrt{S}}{2}\approx 5.00\ GeV$~\cite{Braguta:2006wr,Sun:2009zk}. Here, we take $\mu=5.00\ GeV$ to do our discussion.
\begin{table}[h]
\caption{The form factors and cross sections at different energy scales with the threshold parameters $s_{01}=3.7^{2}\ GeV^{2}$, $s_{02}=4.1^{2}\ GeV^{2}$.}\label{Parameters}
\setlength{\tabcolsep}{11mm}{
\begin{tabular}{ccccc}
\hline\hline
$\mu (\rm GeV)$ &F$_{VP}(\rm GeV^{-1})$   & $\sigma (fb)$   \\ \hline
1.80   & 0.00038$\pm 0.000002$ & 0.26$\pm 0.02$  \\
3.50   & 0.00202$\pm 0.00003$  & 9.12$\pm 0.13$  \\
4.00   & 0.00227$\pm 0.00003$ & 11.30$\pm 0.32$ \\
5.00   & 0.00262$\pm 0.00003$ & 15.60$\pm 0.40$ \\ \hline\hline
\end{tabular}}
\end{table}

We compare the results of experiment and various theoretical methods for this process in TABLE II. The second and the third columns are the results of NRQCD~\cite{Braaten:2002fi}, without and with relativistic correction, respectively. Easy to see, the result is greatly increased after considering the relativistic correction. But in both cases, the next leading order corrections are huge, which reduces the reliability of the results. The fourth column is the result of the light cone expansion method~\cite{Braguta:2005kr,Braguta:2006nf} . Because of the use of relativistic  distribution amplitude of $\psi(2S)$ meson, they get exactly the same result as the experiment. The drawback of this method is that the DA of $\psi(2S)$ are not well known. The fifth column is the result of the method of pQCD and relativistic quark model~\cite{Ebert:2008kj}, which is very close to the experiment. This method takes into account many correction , such as the relativistic $v/c$ corrections to the wave functions, relativistic $\frac{p}{\sqrt{s}}$ corrections emerging from the expansion of the quark and gluon propagators and others. However, this method has the phenomenological structure which will bring about some arbitrariness of phenomenological parameters. The last column, $\sigma=15.60\pm 0.40\ fb$, is our result with a relativistic distribution amplitude of $\psi(2S)$ at the energy scale $\mu=5.00\ GeV$, which is also very close to the experiment data, while the error is caused by Borel parameter $M^{2}$ from $15\ GeV^{2}$ to $25\ GeV^{2}$. From the above, we can see that the relativistic effect is very important for this $e^{+}+e^{-}\rightarrow \psi \left ( 2S \right )+\eta _{c}$ process and the results including the relativistic effects are close to the experiments.
\begin{table}[h]
\caption{Comparison of cross sections of our work with experiment and other methods.}\label{Parameters}
\setlength{\tabcolsep}{4mm}{
\begin{tabular}{cccccccc}
\hline\hline
$\sigma _{Belle}\times Br_{>2}(fb)$~\cite{Abe:2004ww} & $\sigma _{NRQCD}(fb)$~\cite{Braaten:2002fi} & $\sigma(fb)$~\cite{Braaten:2002fi} & $\sigma_{LC}(fb)$~\cite{Braguta:2005kr,Braguta:2006nf} & $\sigma(fb)$~\cite{Ebert:2008kj}   & This work$(fb)$   \\ \hline
16.3$\pm 4.6\pm 3.9$   & 0.96$\pm 0.45$ & 5.6$_{-3.3}^{+10.5}$ & 16.3  & 15.3$\pm 2.4$ & 15.60$\pm 0.40$ \\ \hline\hline
\end{tabular}}
\end{table}

\section{summary}\label{sec5}

In this paper, the form factor $F_{VP}=0.00262\pm 0.00003\ GeV^{-1}$ and cross section $ \sigma=15.60\pm 0.40\ fb$ of the exclusive process $e^{+}+e^{-}\rightarrow \psi \left ( 2S \right )+\eta _{c}$ are obtained within the LCSR with a relativistic distribution amplitude. Combining our previous work~\cite{Sun:2009zk} with our current work, we have succeeded in applying the LCSR to the $e^{+}+e^{-}\rightarrow J/\psi +\eta _{c}$ and $e^{+}+e^{-}\rightarrow \psi \left ( 2S \right )+\eta _{c}$ processes. Because the energy scale involved in Belle experiment is much larger than the initial scale of $\eta_{c}$ used here, we have to consider the evolution of the distribution amplitude from the initial scale to an effective scale, which is taken as $\mu=5.00\ GeV$. Such a result also shows that the relativistic effect is very large for this process.

However, our approach still has limitations. The first point is that we don't know much about the the distribution amplitude of charmonium, so getting a more comprehensive look at the properties of charmonium is for future work. Second, we only consider the contribution of the leading order. We expect that, after considering the contribution of higher order, such as the next leading order correction, the radiative correction, the results will be more reliable. The third point is that our results seem to depend on the energy scale. However, a recent research for the similar exclusive process $e^{+}+e^{-}\rightarrow J/\psi +\eta _{c}$~\cite{Sun:2018rgx} show that, the dependence on the energy scale can in fact be eliminated by applying the principle of maximum conformality (PMC) to set the renormalization scale. By carefully applying the PMC to different topologies of the annihilation process, one can achieves precise prediction. We expect the final result, including all of the corrections and the evolution of the scale, to be independent of the energy scale.

\section*{ACKNOWLEDGMENTS}

This work was supported in part by Natural Science Foundation of China under Grant No.11365018 and No.11375240.

Han-Jie Tao would like to thank Si-Qiang Luo and Fu-Lai Wang for useful discussion.



\begin{thebibliography}{99}
\bibitem{Abe:2002rb}
  K.~Abe {\it et al.} [Belle Collaboration],
  Phys.\ Rev.\ Lett.\  {\bf 89}, 142001 (2002)
  [hep-ex/0205104].
\bibitem{Abe:2004ww}
  K.~Abe {\it et al.} [Belle Collaboration],
  Phys.\ Rev.\ D {\bf 70}, 071102 (2004)
  [hep-ex/0407009].
\bibitem{Braaten:2002fi}
  E.~Braaten and J.~Lee,
  Phys.\ Rev.\ D {\bf 67}, 054007 (2003)
  Erratum: [Phys.\ Rev.\ D {\bf 72}, 099901 (2005)]
  [hep-ph/0211085].
\bibitem{Braguta:2005kr}
  V.~V.~Braguta, A.~K.~Likhoded and A.~V.~Luchinsky,
  Phys.\ Rev.\ D {\bf 72}, 074019 (2005)
  [hep-ph/0507275].
\bibitem{Braguta:2006nf}
  V.~V.~Braguta, A.~K.~Likhoded and A.~V.~Luchinsky,
  Phys.\ Lett.\ B {\bf 635}, 299 (2006)
  [hep-ph/0602047].
\bibitem{Braguta:2012zza}
  V.~V.~Braguta, A.~K.~Likhoded and A.~V.~Luchinsky,
  Phys.\ Atom.\ Nucl.\  {\bf 75}, 97 (2012)
  [Yad.\ Fiz.\  {\bf 75}, 88 (2012)].
\bibitem{Ebert:2006xq}
  D.~Ebert and A.~P.~Martynenko,
  Phys.\ Rev.\ D {\bf 74}, 054008 (2006)
  [hep-ph/0605230].
\bibitem{Martynenko:2013eoa}
  A.~P.~Martynenko and A.~M.~Trunin,
  Phys.\ Rev.\ D {\bf 89}, no. 1, 014004 (2014)
  [arXiv:1308.3998 [hep-ph]].
\bibitem{Ebert:2008kj}
  D.~Ebert, R.~N.~Faustov, V.~O.~Galkin and A.~P.~Martynenko,
  Phys.\ Lett.\ B {\bf 672}, 264 (2009)
  [arXiv:0803.2124 [hep-ph]].
\bibitem{Liu:2004ga}
  K.~Y.~Liu, Z.~G.~He and K.~T.~Chao,
  Phys.\ Rev.\ D {\bf 77}, 014002 (2008)
  [hep-ph/0408141].
\bibitem{Xi-Huai:2014iaa}
  X.~H.~Li and J.~X.~Wang,
  Chin.\ Phys.\ C {\bf 38}, 043101 (2014)
  [arXiv:1301.0376 [hep-ph]].
\bibitem{Shifman:1978bx}
  M.~A.~Shifman, A.~I.~Vainshtein and V.~I.~Zakharov,
  Nucl.\ Phys.\ B {\bf 147}, 385 (1979).
\bibitem{Colangelo:2000dp}
  P.~Colangelo and A.~Khodjamirian,
  In Shifman, M. (ed.): At the frontier of particle physics, vol. 31495-1576
  [hep-ph/0010175].
\bibitem{Sun:2009zk}
  Y.~J.~Sun, X.~G.~Wu, F.~Zuo and T.~Huang,
  Eur.\ Phys.\ J.\ C {\bf 67}, 117 (2010)
  [arXiv:0911.0963 [hep-ph]].
\bibitem{Shifman:1978by}
  M.~A.~Shifman, A.~I.~Vainshtein and V.~I.~Zakharov,
  Nucl.\ Phys.\ B {\bf 147}, 448 (1979).
\bibitem{BHL}
S. J. Brodsky, T. Huang and G. P. Lepage, in Particles and Fields-2, Proceedings of the Banff Summer Institute, Banff, Alberta, 1981, edited by A. Z. Capri and A. N. Kamal (Plenum,
New York, 1983), p143; G. P. Lepage, S. J. Brodsky, T. Huang, and P. B. Mackenize, ibid., p83; T. Huang, in Proceedings of XXth International Conference on High Energy Physics,
Madison, Wisconsin, 1980, edited by L. Durand and L. G. Pondrom, AIP Conf. Proc. No. 69
(AIP, New York, 1981), p1000.
\bibitem{Braguta:2006wr}
  V.~V.~Braguta, A.~K.~Likhoded and A.~V.~Luchinsky,
  Phys.\ Lett.\ B {\bf 646}, 80 (2007)
  [hep-ph/0611021].

\bibitem{Bondar:2004sv}
  A.~E.~Bondar and V.~L.~Chernyak,
  Phys.\ Lett.\ B {\bf 612}, 215 (2005)
  [hep-ph/0412335].
\bibitem{Bodwin:2006dm}
  G.~T.~Bodwin, D.~Kang and J.~Lee,
  Phys.\ Rev.\ D {\bf 74}, 114028 (2006)
  [hep-ph/0603185].
\bibitem{Ma:2004qf}
  J.~P.~Ma and Z.~G.~Si,
  Phys.\ Rev.\ D {\bf 70}, 074007 (2004)
  [hep-ph/0405111].
\bibitem{H}
  Elementary particle Thoery group, Peking University,
   Acta Physica Sinica 25, 316, 415 (1976).
\bibitem{Huang:2007kb}
  T.~Huang and F.~Zuo,
  Eur.\ Phys.\ J.\ C {\bf 51}, 833 (2007)
  [hep-ph/0702147].
\bibitem{Melosh:1974cu}
  H.~J.~Melosh,
  Phys.\ Rev.\ D {\bf 9}, 1095 (1974).
\bibitem{Huang:1994dy}
  T.~Huang, B.~Q.~Ma and Q.~X.~Shen,
  Phys.\ Rev.\ D {\bf 49}, 1490 (1994)
  [hep-ph/9402285].
\bibitem{Tanabashi:2018oca}
  M.~Tanabashi {\it et al.} [Particle Data Group],
  Phys.\ Rev.\ D {\bf 98}, no. 3, 030001 (2018).
\bibitem{Lepage:1980fj}
  G.~P.~Lepage and S.~J.~Brodsky,
  Phys.\ Rev.\ D {\bf 22}, 2157 (1980).
\bibitem{Edwards:2000bb}
  K.~W.~Edwards {\it et al.} [CLEO Collaboration],
  Phys.\ Rev.\ Lett.\  {\bf 86}, 30 (2001)
  [hep-ex/0007012].
\bibitem{Braga:2015jca}
  N.~R.~F.~Braga, M.~A.~Martin Contreras and S.~Diles,
  Phys.\ Lett.\ B {\bf 763}, 203 (2016)
  [arXiv:1507.04708 [hep-th]].
\bibitem{Negash:2015rua}
  H.~Negash and S.~Bhatnagar,
  Int.\ J.\ Mod.\ Phys.\ E {\bf 25}, no. 08, 1650059 (2016)
  [arXiv:1508.06131 [hep-ph]].
\bibitem{Eidemuller:2000rc}
  M.~Eidemuller and M.~Jamin,
  Phys.\ Lett.\ B {\bf 498}, 203 (2001)
  [hep-ph/0010334].
\bibitem{Sun:2018rgx}
  Z.~Sun, X.~G.~Wu, Y.~Ma and S.~J.~Brodsky,
  Phys.\ Rev.\ D {\bf 98}, no. 9, 094001 (2018)
  [arXiv:1807.04503 [hep-ph]].
\end{thebibliography}
\end{document}